\documentclass[12pt]{article}
\usepackage{bbm}
\usepackage{mathrsfs}
\usepackage{amsfonts}
\usepackage{amsmath}
\usepackage{amssymb}
\usepackage{mathptmx}
\usepackage[dvips]{color}
\usepackage{epsfig}
\usepackage{graphicx}
\usepackage{cite}
\usepackage{color}

\def\comment#1{}

\begin{document}
\baselineskip=16pt

\pagenumbering{arabic}

\vspace{1.0cm}

\begin{center}
{\Large\sf Einstein-Euler-Heisenberg theory and charged black holes}
\\[10pt]
\vspace{10pt}

{Remo Ruffini $^{a,b,c}$\footnote{ruffini@icra.it}, Yuan-Bin Wu
$^{a,b,c}$\footnote{Corresponding author. wuyb@icranet.org},
She-Sheng Xue $^{a,b}$\footnote{xue@icra.it}} \vspace{5pt}

{\small
{\it $^a$ Dipartimento di Fisica and ICRA, Sapienza Universit\`a di Roma\\
P.le Aldo Moro 5, I-00185 Rome, Italy
\\
$^b$ ICRANet, P.zza della Repubblica 10, I-65122 Pescara, Italy
\\
$^c$ ICRANet, University of Nice-Sophia Antipolis, 28 Av. de
Valrose, \\
06103 Nice Cedex 2, France} }

\vspace{3.0ex}

{\bf Abstract}

\end{center}

Taking into account the Euler-Heisenberg effective Lagrangian of
one-loop nonperturbative quantum electrodynamics (QED)
contributions, we formulate the Einstein-Euler-Heisenberg theory and
study the solutions of nonrotating black holes with electric and
magnetic charges in spherical geometry. In the limit of strong and
weak electromagnetic fields of black holes, we calculate the black
hole horizon radius, area, and total energy up to the leading order
of QED corrections and discuss the black hole irreducible mass,
entropy, and maximally extractable energy as well as the
Christodoulou-Ruffini mass formula. We find that these black hole
quantities receive the QED corrections, in comparison with their
counterparts in the Reissner-Nordstr\"om solution. The QED
corrections show the screening effect on black hole electric charges
and the paramagnetic effect on black hole magnetic charges. As a
result, the black hole horizon area, irreducible mass, and entropy
increase; however, the black hole total energy and maximally
extractable energy decrease, compared with the Reissner-Nordstr\"om
solution. In addition, we show that the condition for extremely
charged black holes is modified due to the QED correction.

\begin{flushleft}




\end{flushleft}

\newpage

\section{Introduction}
\label{sec:intro}

For several decades the nonlinear electromagnetic generalization of
the Reissner-Nordstr\"om solution of the Einstein-Maxwell equations
has attracted a great deal of attention. The most popular example is
the gravitating Born-Infeld (BI) theory \cite{Born1934}. The static
charged black holes in gravitating nonlinear electrodynamics were
studied in the 1930s \cite{Hoffmann1935, Hoffmann1937}. The
discovery that the string theory, as well as the D-brane physics,
leads to Abelian and non-Abelian BI-like Lagrangians in its
low-energy limit (see, e.g., Refs.~\cite{Fradkin1985,
Abouelsaood1987, Tseytlin1997}), has renewed the interest in these
kinds of nonlinear actions. Asymptotically flat, static, spherically
symmetric black hole solutions for the Einstein-Born-Infeld theory
were obtained in the literature \cite{Garcia1984, Demianski1986}.

Generalization of the exact solutions of spherically symmetric
Born-Infeld black holes with a cosmological constant in arbitrary
dimensions has been considered \cite{Fernando2003, Dey2004,
Cai2004}, as well as in other gravitational backgrounds
\cite{Wiltshire1998, Aiello2004}. Many other models of nonlinear
electrodynamics leading to static and spherically symmetric
structures have been considered in the last decades, such as the
theory with a nonlinear Lagrangian of a general function of the
gauge invariants ($F^{\mu\nu}F_{\mu\nu}$ and $F_{\mu \nu }\tilde
F^{\mu \nu }$) \cite{Alonso20101, Alonso20102, Alonso20111,
Alonso20112} or a logarithmic function of the Maxwell invariant
($F^{\mu\nu}F_{\mu\nu}$) \cite{Soleng1995}, and the theory with a
generalized nonlinear Lagrangian \cite{Oliveira1994} which can lead
to the BI Lagrangian and the weak-field limit of the
Euler-Heisenberg effective Lagrangian \cite{Heisenberg1936}. The
static and spherically symmetric black hole, whose gravity coupled
to the nonlinear electrodynamics of the weak-field limit of the
Euler-Heisenberg effective Lagrangian as a low-energy limit of the
Born-Infeld theory, was studied in Ref.~\cite{Yajima2001}. Some
attempts in the obtention of regular (singularity-free) static and
spherically symmetric black hole solutions in gravitating nonlinear
electrodynamics have been made \cite{Beato1998, Beato1999,
Lombardo2009, Burinskii2002, Dymnikova2004}, and the unusual
properties of these solutions have been discussed in
Refs.~\cite{Novello2000, Bronnikov2001}. Generalization of
spherically symmetric black holes in higher dimension in the theory
with a nonlinear Lagrangian of a function of power of the Maxwell
invariant has been considered in the literature \cite{Hassaine2007,
Hassaine2008, Gonzalez2009, Mazharimousavi2010}. Finally, we mention
that rotating black branes \cite{Dehghani2006, Dehghani2007} and
rotating black strings \cite{Hendi2010} in the Einstein-Born-Infeld
theory have been also considered.

The effective Lagrangian of nonlinear electromagnetic fields has
been formulated for the first time by Heisenberg and Euler using the
Dirac electron-positron theory \cite{Heisenberg1936}. Schwinger
reformulated this nonperturbative one-loop effective Lagrangian
within the quantum electrodynamics (QED) framework
\cite{Schwinger1951}. This effective Lagrangian characterizes the
phenomenon of vacuum polarization. Its imaginary part describes the
probability of the vacuum decay via the electron-positron pair
production. If electric fields are stronger than the critical value
$E_c = m^2 c^3/e\hbar$, the energy of the vacuum can be lowered by
spontaneously creating electron-positron pairs \cite{Heisenberg1936,
Schwinger1951, Sauter1931}. For many decades, both theorists and
experimentalists have been interested in the aspects of the
electron-positron pair production from the QED vacuum and the vacuum
polarization by an external electromagnetic field (see, e.g.,
Refs.~\cite{Ruffini2010, ELI}).

As a fundamental theory, QED gives an elegant description of the
electromagnetic interaction; moreover, it has been experimentally
verified. Therefore, it is important to study the QED effects in
black hole physics. As a result of one-loop nonperturbative QED, the
Euler-Heisenberg effective Lagrangian deserves to attract more
attention in the topic of generalized black hole solutions mentioned
above. In this article, we adopt the contribution from the
Euler-Heisenberg effective Lagrangian to formulate the
Einstein-Euler-Heisenberg theory, and study the solutions of
electrically and magnetically charged black holes in spherical
geometry. We calculate and discuss the QED corrections to the black
hole horizon area, entropy, total energy, and the maximally
extractable energy.

The article is organized as follows. In Sec.~\ref{sec:EH}, we first
recall the Euler-Heisenberg effective Lagrangian. We formulate the
Einstein-Euler-Heisenberg theory in Sec.~\ref{sec:EEHT}. The study
of electrically charged black holes in the weak electric field case
is presented in Sec.~\ref{sec:NRCBH}. The study of magnetically
charged black holes in both weak and strong magnetic field cases is
presented in Sec.~\ref{sec:NRMCBH}. Then we present the study of
black holes with both electric and magnetic charges in the
Einstein-Euler-Heisenberg theory in Sec.~\ref{sec:NREMCBH}. A
summary is given in Sec.~\ref{sec:sum}. The use of units with $\hbar
= c = 1$ is throughout the article.

\section{The Euler-Heisenberg effective Lagrangian}
\label{sec:EH}

The QED one-loop effective Lagrangian was obtained by Heisenberg and
Euler \cite{Heisenberg1936} for constant electromagnetic fields,
\begin{eqnarray} \label{oneloopl}
  \Delta{\mathcal L}_{\rm eff} &=& {1\over 2(2\pi)^2}\int_0^\infty {ds\over s^3}
  \Big[e^2\varepsilon\beta s^2\coth(e\varepsilon s )\cot(e\beta s )\nonumber\\
  & & -1-{e^2\over 3}(\varepsilon^2-\beta^2)s^2\Big]
  e^{-is(m^2_e-i\eta)},
\end{eqnarray}
as a function of two invariants: the scalar $S$ and the pseudoscalar
$P$,
\begin{eqnarray} \label{lightlike}
  S &\equiv & - \frac{1}{4}F_{\mu \nu }F^{\mu \nu }= \frac{1}{2}({\bf E}^2-{\bf B}^2)\equiv \varepsilon^2- \beta^2, \nonumber\\
  P &\equiv & - \frac{1}{4}F_{\mu \nu }\tilde F^{\mu \nu }=
  {\bf E}\cdot {\bf B}\equiv \varepsilon \beta,
\end{eqnarray}
where the field strength is $F^{\mu\nu}$, $\tilde F^{\mu\nu} \equiv
\epsilon^{\mu\nu\lambda\kappa}F_{\lambda\kappa }/2$, and
\begin{eqnarray}
  \varepsilon &=& \sqrt{(S^2+P^2)^{1/2} + S}, \label{fieldinvarianta}\\
  \beta &=& \sqrt{(S^2+P^2)^{1/2}- S}. \label{fieldinvariantb}
\end{eqnarray}
The effective Lagrangian reads
\begin{equation} \label{comEH}
  {\mathcal L}_{\rm eff} = {\mathcal L}_M + \Delta{\mathcal L}_{\rm eff},
\end{equation}
where ${\mathcal L}_M=S$ is the Maxwell Lagrangian. Its imaginary
part is related to the decay rate of the vacuum per unit volume
\cite{Heisenberg1936,Schwinger1951},
\begin{equation} \label{probabilityeh}
  \frac{\Gamma}{V} = \frac{\alpha\varepsilon^2}{\pi^2}
  \sum_{n=1} \frac{1}{n^2} \frac{n\pi\beta/\varepsilon}{\tanh
  {n\pi\beta/\varepsilon}} \exp\left(-\frac{n\pi E_c}{\varepsilon}\right)
\end{equation}
for fermionic fields, and
\begin{equation} \label{probabilityehb}
  \frac{\Gamma}{V} = \frac{\alpha\varepsilon^2}{2\pi^2}
  \sum_{n=1} \frac{(-1)^n}{n^2} \frac{n\pi\beta/\varepsilon}
  {\sinh{n\pi\beta/\varepsilon}}\exp\left(-\frac{n\pi E_c}{
  \varepsilon}\right)
\end{equation}
for bosonic fields; here, $E_c = \frac{m_e^2 c^3}{e\hbar}$ is the
critical field. Using the expressions \cite{Gradshteyn1994}
\begin{eqnarray}
  e\varepsilon s \coth{(e\varepsilon s)} &=&
  \sum_{n=-\infty}^{\infty} \frac{s^2}{(s^2 + \tau_n^2)}, \quad \tau_n
  \equiv n\pi/e\varepsilon,\\
  e\beta s \cot{(e\beta s)} &=& \sum_{m=-\infty}^{\infty} \frac{s^2}{(s^2 -
  \tau_m^2)}, \quad \tau_m \equiv m\pi/e\beta,
\end{eqnarray}
one obtains the real part of the Euler-Heisenberg effective
Lagrangian (\ref{oneloopl}) (see Refs.~ \cite{Ruffini2010,
Ruffini2006, Mielniczuk1982, Valluri1993, Cho2001, ksx2013}),
\begin{eqnarray}
  (\Delta{\mathcal L}^{\cos}_{\rm eff})_{\mathcal P} &=& {1\over2(2\pi)^2}\sum_{n,m=-\infty}^{\infty}
  ~{1\over \tau^2_m+\tau^2_n} \Big[\bar\delta_{m0}J(i\tau_m m^2_e)-\bar\delta_{n0}J(\tau_n
  m^2_e)\Big] \label{pertur}\\
  &=& -\frac{1}{(2\pi)^2} \left[ \sum_{n=1}^{\infty} \frac{e\beta}{\tau_n} \coth{(e\beta \tau_n)} J(\tau_n m_e^2) -
  \sum_{m=1}^{\infty} \frac{e\epsilon}{\tau_m} \coth{(e\epsilon \tau_m)} J(i\tau_m m_e^2)\right]. \label{perturRS}
\end{eqnarray}
The symbol $\bar\delta_{ij} \equiv 1-\delta_{ij}$ denotes the
complimentary Kronecker $\delta$, which vanishes for $i=j$, and
\begin{equation} \label{Jz1}
  J(z) \equiv {\mathcal P}\int^\infty_0 ds {s e^{-s}\over s^2-z^2} =
  -\frac{1}{2}\Big[e^{-z}{\rm Ei}(z) + e^{z}{\rm Ei}(-z)\Big].
\end{equation}
Here, $\mathcal P$ indicates the principle value integral, and ${\rm
Ei}(z)$ is the exponential-integral function,
\begin{equation} \label{Jz2}
  {\rm Ei}(z) \equiv {\mathcal P}\int_{-\infty}^z dt
  \frac{e^t}{t}=\log(-z)+\sum_{k=1}^\infty\frac{z^k}{kk!}.
\end{equation}

Using the series and asymptotic representation of the
exponential-integral function ${\rm Ei}(z)$ for large $z$
corresponding to weak electromagnetic fields ($\varepsilon/E_c\ll
1$, $\beta/E_c\ll 1$),
\begin{equation} \label{Jz3}
  J(z) =-\frac{1}{z^2}-\frac{6}{z^4}-\frac{120}{z^6}
  -\frac{5040}{z^8}-\frac{362880}{z^{10}}+\cdot\cdot\cdot,
\end{equation}
the weak-field expansion of Eq.~(\ref{pertur}) is
\begin{equation} \label{Kleinert1}
  (\Delta{\mathcal L}_{\rm eff})_{\mathcal P} = \frac{2\alpha^2}
  {45 m_e^4}\left\{4S^2+7P^2\right\} + \frac{64\pi\alpha ^3}{315 m_e^8}
  \left\{16S^3+26SP^2\right\}+\cdot\cdot\cdot,
\end{equation}
which is expressed in terms of a powers series of weak
electromagnetic fields up to $O(\alpha^3)$, the first term was
obtained by Heisenberg and Euler in their original article
\cite{Heisenberg1936}.

On the other hand, using the series and asymptotic representation of
the exponential-integral function ${\rm Ei}(z)$ for small $z\ll 1$
\cite{Gradshteyn1994} corresponding to strong electromagnetic fields
($\varepsilon/E_c\gg 1$, $\beta/E_c\gg 1$),
\begin{equation} \label{smallz}
  J(z) = -\frac{1}{2}\Big[e^z\ln(z)+e^{-z}\ln(-z)\Big]-
  \frac{1}{2}\gamma \Big[e^z + e^{-z}\Big]
  +{\mathcal O}(z),
\end{equation}
the leading terms in the strong-field expansion of
Eqs.~(\ref{pertur}) and (\ref{perturRS}) are given by (see
Refs.~\cite{Ruffini2010, Ruffini2006, ksx2013, Kleinert2011})
\begin{eqnarray}
  & & (\Delta{\mathcal L}^{\rm cos}_{\rm eff})_{\mathcal P}
  = {1\over2(2\pi)^2}\sum_{n,m=-\infty}^{\infty} {1\over
  \tau^2_m+\tau^2_n}\Big[\bar\delta_{n0} \ln{(\tau_n m_e^2)}
  -\bar\delta_{m0} \ln{(\tau_m m_e^2)}\Big] + \cdot\cdot\cdot \label{strongexp} \\
  & & \quad = \frac{1}{2(2\pi)^2} \bigg[ \sum_{n=1}^{\infty} \frac{e\beta}{\tau_n} \coth{(e\beta \tau_n)}
  \ln{(\tau_n m_e^2)} - \sum_{m=1}^{\infty} \frac{e\epsilon}{\tau_m} \coth{(e\epsilon \tau_m)}
  \ln{(\tau_m m_e^2)} \bigg] + \cdot\cdot\cdot. \label{strongexpRS}
\end{eqnarray}
In the case of vanishing magnetic field ${\bf B}=0$ and a strong
electric field $E\gg E_c$ using $\lim_{z\rightarrow \infty} J(iz) =
0$ and $\lim_{z\rightarrow 0} z\coth{(az)} = 1/a$,
Eq.~(\ref{strongexpRS}) becomes (see Refs.~\cite{Ruffini2010,
Ruffini2006, ksx2013})
\begin{eqnarray}
  (\Delta{\mathcal L}^{\rm cos}_{\rm eff})_{\mathcal P}
  &=& {e^2E^2\over4\pi^4}\sum_{n=1}^{\infty}
  {1\over n^2} \left[\ln\left(\frac{n\pi E_c}{E}\right) + \gamma
  \right] + \cdot\cdot\cdot \label{strongexpe}\\
  &=& \frac{e^2 E^2}{24\pi^2} \left[ \ln{\left( \frac{\pi E_c}{E} \right)} + \gamma
  \right] - \frac{e^2 E^2}{4\pi^4} \zeta'(2) + \cdot\cdot\cdot, \label{strongexpesum}
\end{eqnarray}
with the Euler-Mascheroni constant $\gamma=0.577216$, the Riemann
zeta function $\zeta(k) = \sum_n 1/n^k$, and
\begin{equation} \label{dRf2}
  \zeta'(2) = \frac{\pi^2}{6} \left[ \gamma + \ln{(2\pi)} - 12 \ln{A}
  \right] \simeq - 0.937548,
\end{equation}
with $A=1.28243$ being the Glaisher constant. Similarly, in the case
of vanishing electric field ${\bf E}=0$ and a strong magnetic field
$B\gg E_c$,  Eq.~(\ref{strongexpRS}) becomes (see
Refs.~\cite{Ruffini2010, Ruffini2006, ksx2013})
\begin{eqnarray}
  (\Delta{\mathcal L}^{\rm cos}_{\rm eff})_{\mathcal P}
  &=& -{e^2B^2\over4\pi^4}\sum_{m=1}^{\infty}
  {1\over n^2} \left[\ln\left(\frac{n\pi E_c}{B}\right) + \gamma
  \right] + \cdot\cdot\cdot \label{strongexpb}\\
  &=& -\frac{e^2 B^2}{24\pi^2} \left[ \ln{\left( \frac{\pi E_c}{B} \right)} + \gamma
  \right] + \frac{e^2 B^2}{4\pi^4} \zeta'(2) + \cdot\cdot\cdot. \label{strongexpbsum}
\end{eqnarray}
The ($n=1$) term in Eq.~(\ref{strongexpb}) is the one obtained by
Weisskopf \cite{Weisskopf1936}.

\section{The Einstein-Euler-Heisenberg theory}
\label{sec:EEHT}

Since the real part of the Euler-Heisenberg effective Lagrangian
$(\Delta{\mathcal L}^{\rm cos}_{\rm eff})_{\mathcal P}$ of
Eq.~(\ref{pertur}) is expressed in terms of Lorentz invariants
$(\varepsilon, \beta)$ or $(S,P)$, the Euler-Heisenberg effective
action in the curve space-time described by metric $g_{\mu\nu}$ can
be written as
\begin{equation} \label{ehactionc}
  {\mathcal S}_{\rm EH} = \int d^4x \sqrt{-g}{\mathcal L}_{\rm EH},\quad
  {\mathcal L}_{\rm EH}=\left[S + (\Delta{\mathcal L}^{\rm cos}_{\rm
  eff})_{\mathcal P}\right].
\end{equation}
The Einstein and Euler-Heisenberg action is then given by
\begin{equation} \label{eehaction}
  {\mathcal S}_{\rm EEH} = -\frac{1}{16\pi G}\int d^4x \sqrt{-g}R +
  {\mathcal S}_{\rm EH},
\end{equation}
where $R$ is the Ricci scalar.

The Einstein field equations are
\begin{equation} \label{eequation}
  G^{\mu\nu}\equiv R^{\mu\nu} - \frac{1}{2} g^{\mu\nu} R=8\pi G
  T^{\mu\nu},
\end{equation}
where the energy-momentum tensor is
\begin{equation} \label{emt}
  T^{\mu\nu} = \frac{2}{\sqrt{-g}}\frac{\delta {\mathcal S}_{\rm
  EH}}{\delta g_{\mu\nu}}.
\end{equation}
The electromagnetic field equations and Bianchi identities are given
by
\begin{equation} \label{fequation}
  {\mathcal D}_\mu P^{\nu\mu}=j^\nu, \quad {\mathcal D}_\mu \tilde
  F^{\mu\nu} = 0,
\end{equation}
and the displacement fields $P^{\nu\mu}, D^i=P^{0i}$, and
$H^i=-\epsilon ^{ijk} P_{jk}$ are defined as
\begin{equation} \label{dfield}
  P^{\mu\nu} = \frac{\delta {\mathcal L}_{\rm EH}}{\delta F_{\mu\nu}},
  \quad  D^i = \frac{\delta {\mathcal L}_{\rm EH}}{\delta E_i},\quad
  H^i=-\frac{\delta {\mathcal L}_{\rm EH}}{\delta B_i}.
\end{equation}
Here, electromagnetic fields are treated as smooth varying fields
over all space generated by external charge currents $j^\mu$ at
infinity.

Using functional derivatives, we obtain
\begin{eqnarray} \label{ide}
  T^{\mu\nu} &=& -g^{\mu\nu}\left[S + (\Delta{\mathcal L}^{\rm
  cos}_{\rm eff})_{\mathcal P}\right]+ 2\left[ \frac{\delta S}{\delta
  g_{\mu\nu}}\frac{\delta {\mathcal L}_{\rm EH}}{\delta S}+
  \frac{\delta P}{\delta g_{\mu\nu}}\frac{\delta {\mathcal L}_{\rm EH}}{\delta P}\right],\nonumber\\
  &=& -g^{\mu\nu}\left[S + (\Delta{\mathcal L}^{\rm cos}_{\rm
  eff})_{\mathcal P}\right]+ 2\left[(1+{\mathcal A}_S) \frac{\delta
  S}{\delta g_{\mu\nu}}+ {\mathcal A}_P \frac{\delta P}{\delta
  g_{\mu\nu}}\right],
\end{eqnarray}
where two invariants are defined as
\begin{equation} \label{dab}
  {\mathcal A}_S \equiv \frac{\delta (\Delta{\mathcal L}^{\rm
  cos}_{\rm eff})_{\mathcal P}}{\delta S};\quad {\mathcal A}_P\equiv
  \frac{\delta (\Delta{\mathcal L}^{\rm cos}_{\rm eff})_{\mathcal
  P}}{\delta P}.
\end{equation}
It is straightforward to obtain
\begin{equation} \label{dsp}
  \frac{\delta S}{\delta g_{\mu\nu}} = \frac{1}{2} F^{\mu}_{\hskip0.2cm
  \lambda} F^{\lambda\nu},\quad \frac{\delta P}{\delta g_{\mu\nu}} =
  F^\mu_{\hskip0.2cm \lambda} \tilde F^{\lambda\nu} = g^{\mu\nu}P,
\end{equation}
and as a result, we rewrite Eq.~(\ref{ide}) as
\begin{eqnarray} \label{ide1}
  T^{\mu\nu} &=& T^{\mu\nu}_M+g^{\mu\nu}\left[{\mathcal
  A}_PP-(\Delta{\mathcal L}^{\rm cos}_{\rm eff})_{\mathcal P}\right]+
  {\mathcal A}_S F^{\mu}_{\hskip0.2cm \lambda} F^{\lambda\nu},\nonumber\\
  &=& T^{\mu\nu}_M(1+{\mathcal A}_S)+g^{\mu\nu}\left[ {\mathcal A}_S
  S+{\mathcal A}_PP -(\Delta{\mathcal L}^{\rm cos}_{\rm
  eff})_{\mathcal P}\right],
\end{eqnarray}
where $T^{\mu\nu}_M=-g^{\mu\nu}S + F^{\mu}_{\hskip0.2cm \lambda}
F^{\lambda\nu}$ is the energy-momentum tensor of the electromagnetic
fields of the linear Maxwell theory. Equation (\ref{ide1}) is in
fact a general result, independent of the explicit form of nonlinear
Lagrangian $(\Delta{\mathcal L}^{\rm cos}_{\rm eff})_{\mathcal P}$.
Equations (\ref{ehactionc})-(\ref{ide1}) in principle give a
complete set of equations for Einstein and Euler-Heisenberg
effective theory, together with total charge ($Q$), angular-momentum
($L$), and energy ($M$) conservations. In this article, adopting the
Euler-Heisenberg effective Lagrangian (\ref{pertur}), we explicitly
calculate invariants ${\mathcal A}_S$ and ${\mathcal A}_P$ of
Eq.~(\ref{dab}) as well as the energy-momentum $T^{\mu\nu}$ of
Eq.~(\ref{ide1}) in the following cases.

It is necessary to point out that in present article, we do not
consider the couplings between photons and gravitons that are also
induced by QED vacuum polarization effects at the level of
one-fermion loop. Drummond and Hathrell obtained the photon
effective action from the lowest term of one-loop vacuum
polarization on a general curved background manifold; i.e., a
graviton couples to two on-mass-shell photons through a fermionic
loop \cite{Drummond1980},
\begin{eqnarray} \label{DHaction}
  {\mathcal S}_{\rm DH} &=& -\frac{\alpha}{720\pi m_e^2}\int d^4x \sqrt{-g}
  \left( 5 R F_{\mu\nu} F^{\mu\nu}
  - 26 R_{\mu\nu} F^{\mu\sigma} F^{\nu}_{\sigma}\right. \nonumber\\
  & & \qquad\qquad\qquad\qquad \left. + 2 R_{\mu\nu\sigma\tau}
  F^{\mu\nu} F^{\sigma\tau} + 24 {\mathcal D}_{\mu} F^{\mu\nu}
  {\mathcal D}_{\sigma} F^{\sigma}_{\nu} \right).  \label{DHaction}
\end{eqnarray}
Further studies of one-loop effective action (\ref{DHaction}) were
made based on the approach of the heat-kernel or ``inverse mass''
expansion \cite{Gilkey1975, Bastianelli2000}, the approach of the
so-called ``derivative expansion" \cite{Barvinsky8590, Gusev2009},
and the consideration of the one-loop one particle irreducible of
one graviton interacting with any number of photons
\cite{Bastianelli2009}. This effective action (\ref{DHaction}) was
used to study the modified photon dispersion relation by a generic
gravitational background \cite{Drummond1980} and the possible
consequences \cite{Latorre1995, Dittrich1998, Shore1996,
Hollowood0708}.

At the level of one-loop quantum corrections of the QED theory in
the presence of gravitational field, the effective Lagrangian
(\ref{DHaction}) should be considered as an addition to the Euler
and Heisenberg effective Lagrangian (\ref{Kleinert1}) in the
weak-field limit. In this article, we try to quantitatively study
the QED corrections in spherically symmetric black holes with mass
$M$ and charge $Q$. In this case, the corrections from the Euler and
Heisenberg effective Lagrangian (\ref{Kleinert1}) must be much
larger than the one from the effective Lagrangian (\ref{DHaction}).
Studying the discussion and result of Ref.~\cite{Drummond1980} for
spherical symmetric black holes, we approximately estimate the ratio
of Eqs.~(\ref{Kleinert1}) and (\ref{DHaction}) around the horizon of
black holes with mass $M$ and charge $Q$. As a result, this ratio is
$\sim 10^{-2} \left( \frac{Q}{M\sqrt{G}} \right)^2
\frac{\alpha}{Gm_e^2}\gg 1$. It is not surprising that the
electromagnetic coupling $e\sim 1/\sqrt{137}$ is much larger than
the effective gravitational counterpart $Gm_e^2\sim 10^{-45}$.
Besides, it is expected that calculations involving both the
Euler-Heisenberg effective Lagrangian (\ref{Kleinert1}) and
Eq.~(\ref{DHaction}) are much more complex and tedious.
Nevertheless, it is interesting to investigate the effect of the
photon-graviton amplitudes on black hole physics.  In this article,
for the sake of simplicity, we first consider only the
Einstein-Euler-Heisenberg action (\ref{eehaction}) as a leading
contribution in order to gain some physical insight into the QED
corrections in black hole physics.

\subsection{${\bf B}=0, {\bf E}\not=0$ or ${\bf E}=0, {\bf B}\not=0$}
\label{sec:AEMFO}

We consider the case of ${\bf B}=0 $ and ${\bf E}\not=0$, namely,
$\beta=P=0$, $\varepsilon =E=|{\bf E}|$, and $S=E^2/2$. ${\mathcal
A}_P=0$ and the effective Lagrangian Eq.~(\ref{pertur}) becomes
\begin{equation} \label{perture}
  (\Delta{\mathcal L}^{\cos}_{\rm eff})_{\mathcal P}  = - {e^2E^2\over
  4\pi^4}\sum_{n=1}^{\infty} {1\over n^2} J(n\pi E_c/E).
\end{equation}
Using
\begin{equation} \label{Jint}
  {\mathcal P}\int^\infty_0 ds {e^{-s}\over (s^2-z^2)} = -
  \frac{1}{2z}\Big[ e^{-z}{\rm Ei}(z) - e^{z}{\rm Ei}(-z)\Big],
\end{equation}
we calculate
\begin{equation} \label{dJz}
  \frac{dJ(z)}{dz^2} = {\mathcal P}\int^\infty_0 ds {s e^{-s}\over
  (s^2-z^2)^2} =\frac{1}{2z^2} - {\mathcal P}\int^\infty_0 ds
  {e^{-s}\over (s^2-z^2)}
\end{equation}
and obtain
\begin{eqnarray} \label{as}
  {\mathcal A}_S &=& -
  {e^2\over 2 \pi^4}\sum_{n=1}^{\infty} {1\over n^2}J(n\pi E_c/E)\nonumber\\
  & & -{e^2\over 4 \pi^2}\zeta(2) +
  {e^2\over 4 \pi}\frac{E_c}{E}\sum_{n=1}^{\infty} {1\over n}
  \tilde J(n\pi E_c/E),
\end{eqnarray}
where
\begin{equation} \label{Jz1t}
  \tilde J(z)  = e^{-z}{\rm Ei}(z) - e^{z}{\rm Ei}(-z).
\end{equation}
Substituting these quantities into Eq.~(\ref{ide1}), we obtain the
expression of the energy-momentum tensor $T^{\mu\nu}(\varepsilon)$.
In the case of ${\bf E}=0$ and ${\bf B}\not=0$, the energy-momentum
tensor $T^{\mu\nu}(\beta)$ can be straightforwardly obtained from
$T^{\mu\nu}(\varepsilon)$ by the discrete duality transformation
$\varepsilon \rightarrow i\beta$, i.e., $|{\bf E}|\rightarrow i|{\bf
B}|$. In principle, using the complete Euler-Heisenberg effective
Lagrangian $(\Delta{\mathcal L}^{\rm cos}_{\rm eff})_{\mathcal P}$
(\ref{pertur}) for arbitrary electromagnetic fields ${\bf E}$ and
${\bf B}$, one can obtain the energy-momentum tensor
$T^{\mu\nu}(\varepsilon,\beta)$ of Eq.~(\ref{ide1}). For the reason
of practical calculations, we consider the cases of weak and strong
fields.

\subsection{Weak- and strong-field cases}
\label{sec:WSF}

In the weak-field case, using Eq.~(\ref{Kleinert1}) and calculating
Eqs.~(\ref{ide})-(\ref{ide1}), we obtain
\begin{eqnarray} \label{dabkn}
  {\mathcal A}_S &=& \frac{2 \alpha^2}{45 m_e^4}( 8S) +\frac{64\pi
  \alpha^3}{315 m_e^8}(48S^2+26P^2 )+\cdot\cdot\cdot ,\nonumber\\
  {\mathcal A}_P &=& \frac{2 \alpha^2}{45 m_e^4}( 14 P ) +\frac{64\pi
  \alpha^3}{315 m_e^8}( 52SP )+\cdot\cdot\cdot ,
\end{eqnarray}
and
\begin{equation} \label{ide2}
  T^{\mu\nu} = T^{\mu\nu}_M\left[1+8\left(\frac{2\alpha^2}{45m_e^4}\right)S\right]
  +g^{\mu\nu}\left(\frac{2\alpha^2}{45m_e^4}\right)\left[4S^2 +7P^2
  \right] +\cdot\cdot\cdot,
\end{equation}
up to the leading order.

In strong-field case $\varepsilon/E_c \gg 1$ and $\beta/E_c \gg 1$
using Eq.~(\ref{strongexp}) and calculating
Eqs.~(\ref{ide})-(\ref{ide1}), we obtain
\begin{eqnarray} \label{astrongexp}
  {\mathcal A}_S &=& {1\over2(2\pi)^2}\frac{2}{\varepsilon^2+\beta^2}
  \sum_{n,m=-\infty}^{\infty} {1\over (\tau^2_m+\tau^2_n)^2}
  \Big\{\bar\delta_{n0}\Big[(\tau^2_n-\tau^2_m)\ln(\tau_n m^2_e)-\frac{1}{2}(\tau^2_m+\tau^2_n)\Big]\nonumber\\
  & & -\bar\delta_{m0}\Big[(\tau^2_n-\tau^2_m)\ln(\tau_m
  m^2_e)+\frac{1}{2}(\tau^2_m+\tau^2_n)\Big]\Big\}+\cdot\cdot\cdot
\end{eqnarray}
and
\begin{eqnarray} \label{pstrongexp}
  {\mathcal
  A}_P &=& {1\over2(2\pi)^2}\frac{2\varepsilon\beta}{\varepsilon^2+\beta^2}
  \sum_{n,m=-\infty}^{\infty} {1\over (\tau^2_m+\tau^2_n)^2}
  \Big\{\bar\delta_{n0}\Big[\left(\frac{\tau^2_n}{\varepsilon^2}
  +\frac{\tau^2_m}{\beta^2}\right)\ln(\tau_n m^2_e)-\frac{1}{2}\frac{(\tau^2_m+\tau^2_n)}{\varepsilon^2}\Big]\nonumber\\
  & & -\bar\delta_{m0}\Big[\left(\frac{\tau^2_n}{\varepsilon^2}+\frac{\tau^2_m}{\beta^2}\right)\ln(\tau_m
  m^2_e)-\frac{1}{2}\frac{(\tau^2_m+\tau^2_n)}{\beta^2}\Big]\Big\}+\cdot\cdot\cdot .
\end{eqnarray}
From Eq.~(\ref{strongexpesum}) for ${\bf B}=0$ and a strong electric
field, we obtain
\begin{equation} \label{astrongsum}
  {\mathcal A}_S
  = {e^2\over24\pi^2} \left[2\ln\left(\frac{\pi
  E_c}{E}\right) + 2\gamma -1 \right]-{e^2\over2\pi^4}\zeta'(2)+\cdot\cdot\cdot,
\end{equation}
and the energy-momentum tensor $T^{\mu\nu}$ of Eq.~(\ref{ide1}),
\begin{equation} \label{emtnresf}
  T^{\mu\nu} = T^{\mu\nu}_M\left\{1+{e^2\over 24
  \pi^2} \left[ 2\ln{\left( \frac{\pi E_c}{E} \right)} + 2\gamma -1 \right]
  - \frac{e^2}{2\pi^4} \zeta'(2) \right\} - g^{\mu\nu} \frac{e^2 E^2}{48\pi^2} +\cdots.
\end{equation}
Analogously, from Eq.~(\ref{strongexpbsum}) for ${\bf E}=0$ and a
strong magnetic field, we obtain
\begin{equation} \label{apstrongsum}
  {\mathcal A}_S
  = {e^2\over24\pi^2} \left[2\ln\left(\frac{\pi
  E_c}{B}\right) + 2\gamma -1 \right]-{e^2\over2\pi^4}\zeta'(2)+\cdot\cdot\cdot,
\end{equation}
and the energy-momentum tensor
\begin{equation} \label{emtnrmsf}
  T^{\mu\nu} = T^{\mu\nu}_M\left\{1+{e^2\over 24
  \pi^2} \left[ 2\ln{\left( \frac{\pi E_c}{B} \right)} + 2\gamma -1 \right]
  - \frac{e^2}{2\pi^4} \zeta'(2) \right\} + g^{\mu\nu} \frac{e^2 B^2}{48\pi^2} +\cdots.
\end{equation}
In the following sections, using the energy-momentum tensors
$T^{\mu\nu}$ of Eqs.~(\ref{ide2}), (\ref{emtnresf}), and
(\ref{emtnrmsf}), we try to study the solutions of the
Einstein-Euler-Heisenberg theory for nonrotating (spherically
symmetric), electrically or magnetically charged black holes.

\section{Electrically charged black holes}
\label{sec:NRCBH}

In this section, we study a nonrotating (spherically symmetric)
electrically charged black hole. In this spherical symmetry case,
the gauge potential is
\begin{equation} \label{gaugep}
  A_\mu(x)= [A_0(r),0,0,0],
\end{equation}
corresponding to the electric field $E(r)=-A'_0(r)=-\partial
A_0(r)/\partial r$ in the radial direction, and the metric field is
assumed to be
\begin{equation} \label{anline}
ds^2 = f(r)dt^2-f(r)^{-1}dr^2-r^2d\Omega;\quad  f(r)\equiv
1-2Gm(r)/r.
\end{equation}
The metric function $f(r)$ and the electric field $E(r)$ fulfill the
Einstein equations (\ref{eequation}) and electromagnetic field
equations (\ref{fequation}) and their asymptotically flat solutions
at $r\gg 1$,
\begin{equation} \label{aso}
  A_0(r)\rightarrow -\frac{Q}{4\pi r},\quad E(r)\rightarrow
  \frac{Q}{4\pi r^2},\quad \frac{Gm(r)}{r}\rightarrow
  \frac{GM}{r}
\end{equation}
satisfy the Gauss law, where $Q$ and $M$ are the black hole electric
charge and mass seen at infinity.

In order to find the solution near to the horizon of the black hole
by taking into account the QED effects, we approximately adopt the
Euler-Heisenberg effective Lagrangian for constant fields that leads
to the energy-momentum tensor (\ref{ide2}) or (\ref{emtnresf}) for
${\bf B}=0$. This approximation is based on the assumption that the
macroscopic electric field $E(r)$ is approximated as a constant
field $E$ over the microscopic scale of the electron Compton
lengths. When the electric field of charged black holes are
overcritical, electron-positron pair productions take place and the
electric field is screened down to its critical value $E_c$ (see
Refs.~\cite{DR1975, Preparata1998, Preparata2003, Ruffini2008}). In
this article, we study the QED effects on electrically charged black
holes with spherical symmetry, whose electric field is much smaller
than the critical field $E_c$. In this weak electric field case
using Eq.~(\ref{ide2}) we obtain the energy-momentum tensor
\begin{equation} \label{emtnrcbhwf}
  T^{\mu\nu} = T_M^{\mu\nu} \left( 1 + \frac{2\alpha E^2}{45\pi E_c^2}
  \right) + g^{\mu\nu} \frac{\alpha E^4}{90\pi E_c^2} + \cdots.
\end{equation}
As a result, the (0-0) component of Einstein equations is
\begin{equation} \label{eeqr0}
  \frac{2m'(r)}{r^2} = 4\pi \left[ E^2(r) + \frac{\alpha}{15\pi}
  E^4(r)/E_c^2\right],
\end{equation}
which relates to the energy conservation. Analogously, using
Eqs.~(\ref{fequation}) and (\ref{dfield}) and the metric of
Eq.~(\ref{anline}), we obtain the field equation up to the leading
order,
\begin{equation} \label{meqr0}
  \frac{2\alpha}{45\pi} E^3(r)/E_c^2 + E(r) =\frac{Q}{4\pi r^2},
\end{equation}
which is the zero component of ${\mathcal D}_\mu P^{\nu\mu}=j^\nu$
of Eq.~(\ref{fequation}) in the spherical symmetry case. This
equation relates to the total charge conservation.

A similar case was studied in Ref.~\cite{Yajima2001}, in which,
however, the effective Lagrangian [the first term in
Eq.~(\ref{Kleinert1})] was considered as a low-energy limit of the
Born-Infeld theory; the coefficients of the $S^2$ and $P^2$ terms in
Eq.~(\ref{Kleinert1}) are treated as free parameters, so as to
either numerically or analytically study the properties of
spherically symmetric black hole solutions in the
Einstein-Euler-Heisenberg system. In the following, in order to
analytically study the QED effects on the black hole solution, we
use the Euler-Heisenberg effective Lagrangian (\ref{Kleinert1}) and
find the black hole solution by a series expansion in powers of
$\alpha$. Introducing $\overline{E}(r)\equiv E(r)/E_c$, up to the
first order of $\alpha$, the solution to Eq.~(\ref{meqr0}) is
approximately given by
\begin{equation} \label{esol1}
  \overline{E}(r) = E_Q \left( 1 - \frac{2\alpha}{45\pi} E^2_Q + \cdots\right),
\end{equation}
where $E_Q \equiv E_Q(r) \equiv Q/(4\pi r^2E_c)$. We find that the
electric field $E(r)$ is smaller than $Q/4\pi r^2$, due to the
charge screening effect of the vacuum polarization. Substituting
this solution (\ref{esol1}) into the Einstein equation
(\ref{eeqr0}), we obtain the integration
\begin{equation} \label{intm}
  m(r) = M-\int_r^\infty 4\pi r^2 dr \frac{1}{2}\left[E^2(r) +
  \frac{\alpha}{15\pi} E^4(r)/E_c^2\right].
\end{equation}
This equation clearly shows that the energy-mass function $m(r)$ of
Eq.~(\ref{anline}) is the total gravitational mass $M$ (attractive)
``screened down'' by the electromagnetic energy (repulsive). In the
Maxwell theory $(\Delta{\mathcal L}^{\cos}_{\rm eff})_{\mathcal P} =
0$ and $E(r) = Q/(4\pi r^2)$, we obtain the Reissner-Nordstr\"om
solution $m(r) = M-Q^2/8\pi r$. In the Euler-Heisenberg system, it
is not proper to make the integration in Eq.~(\ref{intm}), since the
integrand comes from the Euler-Heisenberg effective Lagrangian,
which is valid only for constant fields. In order to gain some
physical insight into the energy-mass function (\ref{intm}), we
integrate Eq.~(\ref{intm}) to the leading order of $\alpha$,
\begin{equation} \label{intma}
  m(r)\approx M - \frac{Q^2}{8\pi r}\left[1-
  \frac{\alpha}{225\pi}\frac{1}{(4\pi)^2}\frac{Q^2}{r^4}\frac{1}{E_c^2}
  \right]=M - \frac{Q^2}{8\pi r}\left[1-
  \frac{\alpha}{225\pi}E_Q^2\right],
\end{equation}
which shows the QED correction to the Reissner-Nordstr\"om solution.
Due to the QED vacuum polarization effect, the black hole charge $Q$
is screened
\begin{equation} \label{screen}
Q\rightarrow Q\left[1-\frac{\alpha}{225\pi}E_Q^2\right]^{1/2}.
\end{equation}
As a consequence, the electrostatic energy of Eq.~(\ref{intma}) is
smaller than $Q^2/(8\pi r)$ in the Reissner-Nordstr\"om solution.

Moreover, we study the QED correction to the black hole horizon. For
this purpose, we define the horizon radius $r_{_H}$ at which the
function $f(r)$ of Eq.~(\ref{anline}) vanishes, i.e., $f(r_{_H})=0$,
leading to
\begin{equation} \label{hri}
\frac{Gm(r_H)}{r_{_H}}= \frac{1}{2}.
\end{equation}
Using the energy-mass function $m(r)$ of Eq.~(\ref{intma}), we
obtain
\begin{equation}
  \frac{GM}{r_{_H}} - \frac{GQ^2}{8\pi r_{_H}^2} \left[ 1 - \frac{\alpha}{225\pi} E_{Qh}^2
  \right] = \frac{1}{2},
\end{equation}
where $E_{Qh} \equiv E_Q(r_{_H})$. Up to the leading order of
$\alpha$, we obtain
\begin{eqnarray}
  r_{_{H+}} &=& GM + \sqrt{G^2 M^2 - \frac{G Q^2}{4\pi} \left[ 1 -
  \frac{\alpha}{225\pi} E_{Q+}^2
  \right]}, \label{rpnrwf}\\
  r_{_{H-}} &=& GM - \sqrt{G^2 M^2 - \frac{G Q^2}{4\pi} \left[ 1 -
  \frac{\alpha}{225\pi} E_{Q-}^2
  \right]}, \label{rnnrwf}
\end{eqnarray}
where $E_{Q+} \equiv E_Q(r_{_{H+}})$ and $E_{Q-} \equiv
E_Q(r_{_{H-}})$. Equation (\ref{rpnrwf}) shows that the black hole
horizon radius $r_{_{H+}}$ becomes larger than the
Reissner-Nordstr\"om one $r_+$ given by Eq.~(\ref{rpnrwf}) for
setting $\alpha=0$. The black hole horizon area $4\pi r_{_{H+}}^2 $
becomes larger than the Reissner-Nordstr\"om one $4\pi r_+^2$ given
by Eq.~(\ref{rpnrwf}) for setting $\alpha=0$. This is again due to
the black hole charge $Q$ screened by the QED vacuum polarization
(\ref{screen}).

In the Reissner-Nordstr\"om solution, the extreme black hole
solution is given by $r_+=r_-$ or $4\pi GM^2=Q^2$. In our case, this
is given by $r_{_{H+}}=r_{_{H-}} = r_{_{H}}$ yielding
\begin{equation} \label{enrbhwf}
  G^2 M^2 - \frac{G Q^2}{4\pi} \left[ 1 - \frac{\alpha}{225\pi} E_{Qh}^2
  \right] = 0.
\end{equation}
From Eqs.~(\ref{rpnrwf}) and (\ref{rnnrwf}), we obtain
\begin{eqnarray}
  4\pi r_{_H}^2 &=& 4\pi G^2 M^2 = GQ^2 \left[ 1 - \frac{\alpha}{225\pi} E_{Qh}^2
  \right]\nonumber\\
  & =& GQ^2 \left[ 1 - \frac{\alpha}{225\pi} \frac{1}{G^2 Q^2
  E_c^2} \right], \label{enrbhwfs}\\
  r_{_{H_\pm}}&\approx & Q \left[ 1 - \frac{\alpha}{225\pi} E_{Qh}^2
  \right]^{1/2} = Q\left[1-\frac{\alpha}{225\pi}
  \frac{1}{(E_cQ)^2}\right]^{1/2}. \label{enrbhwfr}
\end{eqnarray}
In Eq.~(\ref{enrbhwfr}) we adopt $G/4\pi=1$. Due to the QED
correction, the condition of extremely electrically charged black
holes with spherical symmetry changes from $M=Q/4\pi$ to
\begin{eqnarray}
M=\frac{Q}{4\pi}\left[1-\frac{\alpha}{225\pi}
\frac{1}{(E_cQ)^2}\right]^{1/2}. \label{excondi}
\end{eqnarray}
This implies that for a given $M$, the black holes are allowed to
carry more charge $Q$ than the Reissner-Nordstr\"om case. These
results show that when the black hole mass $M$ is fixed, the horizon
area and radius of the extremely electrically charged black hole are
the same as the extreme Reissner-Nordstr\"om one. However, when the
black hole charge $Q$ is fixed, the black hole horizon area and
radius are smaller than those of the extreme Reissner-Nordstr\"om
black hole. The reason is that the charge screening effect decreases
the electrostatic energy; hence, this leads to a smaller mass $M$
for the extreme black hole.

Now we turn to the maximal energy extractable from a black hole. As
pointed out in Ref.~\cite{Christodoulou1971}, the surface area $S_a$
of the black hole horizon is related to the irreducible mass
$M_{ir}$ of the black hole
\begin{equation} \label{mirnrwf}
  S_a = 16\pi G^2 M_{ir}^2 = 4\pi r_{_{H+}}^2,
\end{equation}
where $r_{_{H+}}$ is given by Eq.~(\ref{rpnrwf}). The surface area
of the black hole horizon cannot be decreased by classical processes
\cite{Christodoulou1971, Christodoulou1970, Hawking1971}. Any
transformation of the black hole which leaves fixed the irreducible
mass is called reversible \cite{Christodoulou1971,
Christodoulou1970}. Any transformation of the black hole which
increases its irreducible mass, for instance, the capture of a
particle with nonzero radial momentum at the horizon, is called
irreversible. In irreversible transformations there is always some
kinetic energy that is irretrievably lost behind the horizon. Note
that transformations which arbitrarily close to reversible ones are
the most efficient transformations for extracting energy from a
black hole \cite{Christodoulou1971, Christodoulou1970}. Following
the same argument presented in Ref.~\cite{Christodoulou1971}, and
including the leading-order QED correction (\ref{intma}), we obtain
the Christodoulou-Ruffini mass formula
\begin{equation} \label{crmnrwf}
  M = M_{ir} + \frac{Q^2}{16\pi GM_{ir}} \left[1-
  \frac{\alpha}{225\pi}E_{Q+}^2\right],
\end{equation}
where the electrostatic energy of the black hole is reduced for the
reason that the black hole charge is screened down by the QED vacuum
polarization effect (\ref{screen}).

The properties of the surface area $S_a$ of the black hole horizon
and irreducible mass $M_{ir}$ can also been understood from the
concepts of information theory \cite{Bekenstein1973}. The black hole
entropy $S_{en}$ is introduced as the measure of information about a
black hole interior which is inaccessible to an exterior observer
and is proportional to the surface area $S_a$ of the black hole
horizon \cite{Bekenstein1973}
\begin{equation} \label{entropynrwf}
  S_{en} = S_a/4 = \pi r_{_{H+}}^2.
\end{equation}
The physical content of the concept of the black hole entropy
derives from the generalized second law of thermodynamics: when
common entropy in the black hole exterior plus the black hole
entropy never decreases \cite{Bekenstein1973}. In the
Einstein-Euler-Heisenberg theory, the black hole irreducible mass of
Eq.~(\ref{mirnrwf}) and entropy of Eq.~(\ref{entropynrwf}) with the
QED correction are determined by the horizon radius $r_{_{H+}}$ of
Eq.~(\ref{rpnrwf}) for charged black holes and Eq.~(\ref{enrbhwfs})
for extreme black holes.

Now we consider the physical interpretation of the electromagnetic
term in Eq.~(\ref{crmnrwf}). This term represents the maximal energy
extractable from a black hole, which can be obtained by evaluating
the conserved Killing integral \cite{Ruffini2010, Ruffini2002}
\begin{equation} \label{maxkillint}
  \int_{\Sigma_{t}^{+}} \xi_{+}^{\mu} T_{\mu\nu} d\Sigma^{\nu}
  = 4\pi \int_{r_{_{H+}}}^{\infty} r^2 T_0^0 dr,
\end{equation}
where $\Sigma_{t}^{+}$ is the spacelike hypersurface in the
space-time region that is outside the horizon $r>r_{_{H+}}$
described by the equation $t= {\rm constant}$, with $d\Sigma^{\nu}$
as its surface element vector. $\xi_{+}^{\mu}$ is the static Killing
vector field. This electromagnetic term in Eq.~(\ref{crmnrwf}) is
the total energy of the electromagnetic field and includes its own
gravitational binding energy. Using the energy-momentum tensor of
Eq.~(\ref{emtnrcbhwf}) and weak-field solution (\ref{esol1}), we
obtain the maximal energy extractable from an electrically charged
black hole
\begin{equation} \label{eenrc}
  \varepsilon_{ex} = \frac{Q^2}{8\pi r_{_{H+}}}\left[1-
  \frac{\alpha}{225\pi}E_{Q+}^2\right].
\end{equation}
This shows that the black hole maximal extractable energy decreases
in comparison with the Reissner-Nordstr\"om case ($Q^2/8\pi r_{+}$).
This can be explained by the following: (i) the charge screening
effect decreases the electrostatic energy; (ii) the black hole
horizon radius $r_{_{H+}}$ of Eq.~(\ref{rpnrwf}) increases, leading
to the decrease of the maximally extractable energy, because the
most efficient transformations that extract energy from a black hole
occur near the horizon. For the extremely electrically charged black
hole, the maximally extractable energy is the same as that in the
Reissner-Nordstr\"om case, when the black hole mass $M$ is fixed;
however, it becomes smaller than the Reissner-Nordstr\"om one when
the black hole electric charge $Q$ is fixed.

\section{Magnetically charged black holes}
\label{sec:NRMCBH}

Now we turn to study the Einstein-Euler-Heisenberg theory
(\ref{ide2}) and (\ref{emtnrmsf}) in the presence of the magnetic
field $\bf{B}$. As shown by Eq.~(\ref{probabilityeh}), the magnetic
field $\bf{B}$ does not contribute to the pair-production rate so
that the process of the electron-positron pair production does not
occur for a strong magnetic field $\bf{B}$. For this reason, we
consider black holes with strong magnetic fields. The conventional
black hole with electric and magnetic fields is the rotating charged
black hole of the Kerr-Newman black hole \cite{Newman1965}. However,
the solution to a rotating charged black hole in the
Einstein-Euler-Heisenberg theory is rather complicated, and we do
not consider it in this work. For the sake of simplicity, we study
the nonrotating magnetically charged black hole with spherical
symmetry in order to investigate the QED corrections in the presence
of the magnetic field $\bf{B}$ in the Einstein-Euler-Heisenberg
theory.

For a nonrotating magnetically charged black hole with magnetic
charge $Q_m$, the tensor $F_{\mu\nu}$ compatible with spherical
symmetry can involve only a radial magnetic field $F_{23} = -
F_{32}$. In the Einstein-Maxwell theory, the field equations
(\ref{fequation}) give (see, e.g., Refs.~\cite{Hawking1995,
Gibbons1995})
\begin{equation} \label{brqm}
  F_{23}= \frac{Q_m \sin{\theta}}{4\pi},
\end{equation}
and the gauge potential will be (see, e.g.,
Refs.~\cite{Hawking1995})
\begin{equation} \label{gaugepm}
  A_\mu(x)= [0,0,0,Q_m (1-\cos{\theta})/4\pi].
\end{equation}
The metric is similar to the one of nonrotating electrically charged
black holes,
\begin{equation} \label{mrtricnrmcbh}
  ds^2 = f(r)dt^2-f(r)^{-1}dr^2-r^2d\Omega, \quad f(r)\equiv 1-2Gm(r)/r,
\end{equation}
where $m(r)$ is the mass-energy function. In the Einstein-Maxwell
theory, the metric function $f(r)$ of magnetically charged black
holes with spherical symmetry is given by (see, e.g.,
Refs.~\cite{Hawking1995})
\begin{equation} \label{metricfnrmem}
  f(r) = 1 - \frac{2G M}{r} + \frac{GQ_m^2}{4\pi r^2},
\end{equation}
where $M$ is the black hole mass seen at infinity.

\subsection{Weak magnetic field case}
\label{sec:NRMCBHWF}

Using Eq.~(\ref{ide2}), we obtain the energy-momentum tensor for the
weak magnetic field $B$ case,
\begin{equation} \label{emtnrmcbhwf}
  T^{\mu\nu} = T_M^{\mu\nu} \left( 1 - \frac{2\alpha B^2}{45\pi E_c^2}
  \right) + g^{\mu\nu} \frac{\alpha B^4}{90\pi E_c^2} + \cdots.
\end{equation}
Similar to the analysis of electrically charged black holes with
spherical symmetry, we obtain the (0-0) component of Einstein
equations,
\begin{equation} \label{eeqnrmcwf}
  \frac{2m'(r)}{r^2} = 4\pi \left[ B^2(r) - \frac{\alpha}{45\pi}
  B^4(r)/E_c^2\right].
\end{equation}
For the magnetically charged black hole with spherical symmetry,
only a radial magnetic field is present. The field equations
(\ref{fequation}) give $B(r) = Q_m/(4\pi r^2)$ (see, e.g.,
Refs.~\cite{Yajima2001, Bronnikov2001}). Substituting $B(r)$ into
the Einstein equation (\ref{eeqnrmcwf}), we obtain the mass-energy
function
\begin{equation} \label{eeqnrmciwf}
  m(r) = M-\int_r^\infty 4\pi r^2 dr \frac{1}{2}\left[B^2(r) -
  \frac{\alpha}{45\pi} B^4(r)/E_c^2\right].
\end{equation}
Neglecting the QED correction of the Euler-Heisenberg effective
Lagrangian, Eq.~(\ref{eeqnrmciwf}) gives $m(r) = M-Q_m^2/8\pi r$,
which is the solution of the magnetically charged
Reissner-Nordstr\"om  black hole in the Einstein-Maxwell theory.
Making the integration in Eq.~(\ref{eeqnrmciwf}), one obtains
\cite{Yajima2001}
\begin{equation} \label{mnrmcbhwf}
  m(r)= M - \frac{Q_m^2}{8\pi r}\left[1-
  \frac{\alpha}{225\pi}\frac{1}{(4\pi)^2}\frac{Q_m^2}{r^4}\frac{1}{E_c^2}
  \right]=M - \frac{Q_m^2}{8\pi r}\left[1-
  \frac{\alpha}{225\pi}B_Q^2\right],
\end{equation}
where $B_Q \equiv B_Q(r) \equiv Q_m/(4\pi r^2 E_c)$. As shown in
Eq.~(\ref{mnrmcbhwf}), taking into account the QED vacuum
polarization effect, the total magnetostatic energy is smaller than
$Q_m^2/8\pi r$ in the magnetically charged Reissner-Nordstr\"om
case. This can be understood as follows. In the magnetic field ${\bf
B}$ of the black holes, the vacuum polarization effect results in a
positive magnetic polarization $\bf{M}$. Then the magnetic ${\bf H}$
field defined $\bf{B}=\bf{H}+\bf{M}$ is smaller than the magnetic
field $\bf{B}$. The magnetostatic energy density $\varepsilon_{EM}
\propto \bf{B}\cdot \bf{H}$ decreases. This shows that in weak
magnetic fields, the vacuum polarization effect exhibits the
paramagnetic property.

Compared to the result of the electrically charged black hole in the
first order of $\alpha$, Eqs.~(\ref{intma}) and (\ref{mnrmcbhwf})
have the same expression. One can obtain Eq.~(\ref{mnrmcbhwf}) by
simply replacing $E_Q$ in Eq.~(\ref{intma}) by $B_Q$, namely,
replacing $Q$ by $Q_m$ because of the duality symmetry (see, e.g.,
Ref.~\cite{Hawking1995}). Similar to the analysis of electric
charged black holes, we obtain the horizon radii $r_{_{H_+}}$ and
$r_{_{H_-}}$ of the magnetically charged black hole, up to the
leading order of $\alpha$,
\begin{eqnarray}
  r_{_{H+}} &=& GM + \sqrt{G^2 M^2 - \frac{G Q_m^2}{4\pi} \left[ 1 -
  \frac{\alpha}{225\pi} B_{Q+}^2
  \right]}, \label{rpnrmwf}\\
  r_{_{H-}}  &=& GM - \sqrt{G^2 M^2 - \frac{G Q_m^2}{4\pi} \left[ 1 -
  \frac{\alpha}{225\pi} B_{Q-}^2
  \right]}, \label{rnnrmwf}
\end{eqnarray}
where $B_{Q+} \equiv B_Q(r_{_{H+}})$ and $B_{Q-} \equiv
B_Q(r_{_{H-}})$. The result (\ref{rpnrmwf}) shows that the black
hole horizon radius $r_{_{H+}}$ increases in comparison with the
magnetically charged Reissner-Nordstr\"om one $r_{+}$. This is again
due to the paramagnetic effect of the vacuum polarization that
decreases the magnetostatic energy of the black hole.

Now we turn to the extreme black hole ($  r_{_{H+}}=
r_{_{H-}}=r_{_{H}}$). Similarly, we have
\begin{equation} \label{enrmbhwf}
  G^2 M^2 - \frac{G Q_m^2}{4\pi} \left[ 1 - \frac{\alpha}{225\pi} B_{Qh}^2
  \right] = 0,
\end{equation}
where $B_{Qh} \equiv B_Q(r_{_H})$, and we obtain the black hole
horizon area and radius
\begin{eqnarray}
  4\pi r_{_H}^2 &=& 4\pi G^2 M^2 = GQ_m^2 \left[ 1 - \frac{\alpha}{225\pi} B_{Qh}^2
  \right] = GQ_m^2 \left[ 1 - \frac{\alpha}{225\pi} \frac{1}{G^2 Q_m^2
  E_c^2} \right], \label{enrmbhwfs}\\
  r_{_H}&\approx & Q_m \left[ 1 - \frac{\alpha}{225\pi} B_{Qh}^2
  \right]^{1/2} = Q_m\left[1-\frac{\alpha}{225\pi}
  \frac{1}{(E_cQ_m)^2}\right]^{1/2}.
\end{eqnarray}
In the second line, we adopt $G/4\pi =1$. The QED correction changes
the condition of extremely magnetically charged black holes with
spherical symmetry from $M=Q_m/4\pi$ to
\begin{eqnarray}
  M=\frac{Q_m}{4\pi}\left[1-\frac{\alpha}{225\pi}
  \frac{1}{(E_cQ_m)^2}\right]^{1/2}. \label{excondim}
\end{eqnarray}
The properties of the horizon area and radius of the extremely
magnetically charged black hole are the same as their counterparts
in the extremely electrically charged black hole, given by the
duality transformation $Q \leftrightarrow Q_m$.

Following the same argument presented in
Ref.~\cite{Christodoulou1971}, we obtain the Christodoulou-Ruffini
mass formula
\begin{equation} \label{crmnrmwf}
  M = M_{ir} + \frac{Q_m^2}{16\pi GM_{ir}} \left[1-
  \frac{\alpha}{225\pi}B_{Q+}^2\right]
\end{equation}
for magnetically charged black holes with spherical symmetry in the
Einstein-Euler-Heisenberg theory. One is able to obtain the
irreducible mass $M_{ir}$ by substituting Eq.~(\ref{rpnrmwf}) into
Eq.~(\ref{mirnrwf}), and the black hole entropy $S_{en}$ by
substituting Eq.~(\ref{rpnrmwf}) into Eq.~(\ref{entropynrwf}). The
irreducible mass $M_{ir}$ and the black hole entropy $S_{en}$ in
terms of black hole horizon radius $r_{_{H+}}$ Eq.~(\ref{rpnrmwf})
have the same paramagnetic property in the presence of the QED
vacuum polarization effect, as already discussed.

As shown in Eq.~(\ref{crmnrmwf}), the maximal energy extractable
from a magnetically charged black hole is
\begin{equation} \label{eenrmcwf}
  \varepsilon_{ex} = \frac{Q_m^2}{8\pi r_{_{H+}}}\left[1-
  \frac{\alpha}{225\pi}B_{Q+}^2\right],
\end{equation}
where $r_{_{H+}}$ is given by Eq.~(\ref{rpnrmwf}). The result shows
that the maximal energy extractable from a magnetically charged
black hole is smaller than $\frac{Q_m^2}{8\pi r_{+}}$ of the
magnetically charged Reissner-Nordstr\"om black hole. The reasons
are the following: (i) the vacuum polarization effect decreases the
magnetostatic energy; (ii) the black hole horizon radius $r_{_{H+}}$
of Eq.~(\ref{rpnrmwf}) increases, therefore the maximally
extractable energy decreases. The maximal energy extractable from an
extremely magnetically charged black hole is the same as that from
an extremely magnetically charged Reissner-Nordstr\"om black hole
when the black hole mass $M$ is fixed, while it decreases when the
black hole magnetic charge $Q_m$ is fixed, as we have already
discussed at the end of Sec.~\ref{sec:NRCBH} for the case of the
extremely electrically charged black hole.

\subsection{Strong magnetic field case}
\label{sec:NRMCBHSF}

In this section, we study the magnetically charged black holes with
a strong magnetic field $B(r)$. From Eq.~(\ref{emtnrmsf}), we obtain
the energy-momentum tensor of the magnetically charged black hole
with spherical symmetry in the strong magnetic field case. Analogous
to the weak magnetic field case of magnetically charged black holes
with spherical symmetry, we obtain the (0-0) component of Einstein
equations
\begin{equation} \label{eeqnrmcsf}
  \frac{2m'(r)}{4\pi r^2} = 4\pi\left\{B^2(r) + \frac{e^2 B^2}{12\pi^2}
  \left[ \ln{\left( \frac{\pi E_c}{B} \right)} + \gamma - \frac{6}{\pi^2}\zeta'(2) \right]
  \right\},
\end{equation}
and the field equations (\ref{fequation}) give $B(r) = Q_m/(4\pi
r^2)$. Substituting this magnetic field $B(r)$ into the Einstein
equation (\ref{eeqnrmcsf}), we obtain
\begin{eqnarray}
  m(r) &\approx & M-\int_r^\infty 4\pi r^2 dr \frac{1}{2} \left\{B^2
  + \frac{e^2 B^2}{12\pi^2} \left[ \ln{\left( \frac{\pi E_c}{B} \right)}
  + \gamma  - \frac{6}{\pi^2} \zeta'(2) \right] \right\} \label{eeqnrmcisf}\\
  &\approx & M - \frac{Q_m^2}{8\pi r}\left\{1+ \frac{\alpha}{3\pi}
  \left[ \ln{\left( \frac{\pi}{B_Q} \right)} + \gamma + 2 - \frac{6}{\pi^2} \zeta'(2) \right]
  \right\}. \label{mnrmcbhsf}
\end{eqnarray}
This result is valid for $B\gg E_c$, for which the value of
$\ln{\left( \pi/B_Q \right)} + \gamma + 2 - \frac{6}{\pi^2}
\zeta'(2)$ is negative. As a result, Eq.~(\ref{mnrmcbhsf}) shows
that the total magnetostatic energy in the presence of the vacuum
polarization is smaller than $Q_m^2/8\pi r$ of the magnetically
charged Reissner-Nordstr\"om black hole. Similar to the weak-field
case, this is again due to the paramagnetic effect of the vacuum
polarization that decreases the magnetostatic energy of black holes.
In the strong magnetic field case, the QED vacuum polarization
effect  is much larger than the result (\ref{mnrmcbhwf}) in the
weak-field case, where the QED correction term in
Eq.~(\ref{mnrmcbhwf}) is small for the smallness of
$\alpha/(225\pi)$ and $B_Q^2$. This result (\ref{mnrmcbhsf}) shows a
significant QED effect of the vacuum polarization on the energy of
magnetically charged black holes in the strong magnetic field case.

Now we turn to the study of the black hole horizon radius and area
in the strong magnetic field case. Using the condition $f(r_{_H}) =
0$, we obtain the horizon radii $r_{_{H+}}$ and $r_{_{H-}}$ up to
the leading order of $\alpha$,
\begin{eqnarray}
  r_{_{H+}} &=& GM + \sqrt{G^2 M^2 - \frac{G Q_m^2}{4\pi} \left[ 1+ \frac{\alpha}{3\pi} \mathcal{K}_{NR+}
  \right]}, \label{rpnrmsf}\\
  r_{_{H-}} &=& GM - \sqrt{G^2 M^2 - \frac{G Q_m^2}{4\pi} \left[ 1+ \frac{\alpha}{3\pi} \mathcal{K}_{NR-}
  \right]}, \label{rnnrmsf}
\end{eqnarray}
where
\begin{eqnarray}
  \mathcal{K}_{NR+} &=& \ln{\left( \frac{\pi}{B_{Q+}} \right)} + \gamma +2 - \frac{6}{\pi^2}
  \zeta'(2),\\
  \mathcal{K}_{NR-} &=& \ln{\left( \frac{\pi}{B_{Q-}} \right)} + \gamma +2 - \frac{6}{\pi^2}
  \zeta'(2).
\end{eqnarray}
Equation (\ref{rpnrmsf}) shows that the horizon radius $r_{_{H+}}$
increases in comparison with the magnetically charged
Reissner-Nordstr\"om one $r_{+}$. This is again due to the
paramagnetic effect of the vacuum polarization that decreases the
magnetostatic contribution to the total energy of black holes.

For the case of the extreme black hole
($r_{_{H+}}=r_{_{H-}}=r_{_{H}}$), we have
\begin{equation} \label{enrmbhsf}
  G^2 M^2 - \frac{G Q_m^2}{4\pi} \left[ 1+ \frac{\alpha}{3\pi} \mathcal{K}_{NR} \right] = 0,
\end{equation}
where
\begin{equation}
  \mathcal{K}_{NR} = \ln{\left( \frac{\pi}{B_{Qh}} \right)} + \gamma +2 - \frac{6}{\pi^2}
  \zeta'(2).
\end{equation}
As a result, we obtain
\begin{eqnarray}
  4\pi r_{_H}^2 &=& 4\pi G^2 M^2 = GQ_m^2 \left[1 + \frac{\alpha}{3\pi} \mathcal{K}_{NR} \right], \label{enrmbhsfs} \\
  r_{_H} &\approx & Q_m \left[ 1 + \frac{\alpha}{3\pi} \mathcal{K}_{NR}
  \right]^{1/2}. \label{enrmbhsfr}
\end{eqnarray}
Similar to the weak magnetic field case, the QED correction changes
the condition of extremely magnetically charged black holes with
spherical symmetry from $M=Q_m/4\pi$ to
\begin{eqnarray}
  M=\frac{Q_m}{4\pi}\left[1 + \frac{\alpha}{3\pi} \mathcal{K}_{NR} \right]^{1/2}. \label{excondimsf}
\end{eqnarray}
These results show that the horizon area and radius of the extreme
black hole are the same as their counterparts of the extremely
magnetically charged Reissner-Nordstr\"om black hole, when the black
hole mass $M$ is fixed. Whereas, the black hole magnetic charge
$Q_m$ is fixed, Eqs.~(\ref{enrmbhsfs}) and (\ref{enrmbhsfr}) show
that the black hole horizon area and radius become smaller than
their counterparts of extremely magnetically charged
Reissner-Nordstr\"om black holes. We have discussed this behavior in
Eqs.~(\ref{enrbhwf})-(\ref{excondi}) for the case of extremely
electrically charged black holes.

Analogously, we obtain the Christodoulou-Ruffini mass formula in the
strong-field case of magnetically charged black holes,
\begin{equation} \label{crmnrmsf}
  M = M_{ir} + \frac{Q^2}{16\pi GM_{ir}} \left[1+ \frac{\alpha}{3\pi} \mathcal{K}_{NR+}
  \right].
\end{equation}
It is straightforward to obtain irreducible mass $M_{ir}$  by
substituting Eq.~(\ref{rpnrmsf}) into Eq.~(\ref{mirnrwf}), and the
black hole entropy $S_{en}$ by substituting Eq.~(\ref{rpnrmsf}) into
Eq.~(\ref{entropynrwf}). Analogous to the case of the electrically
charged black hole, the black hole irreducible mass $M_{ir}$ and
entropy $S_{en}$ in the strong magnetic field case depend on the
black hole horizon radius $r_{_{H+}}$ of Eqs.~(\ref{rpnrmsf}) and
(\ref{enrmbhsfs}). Equation (\ref{crmnrmsf}) indicates that the
maximal energy extractable from a magnetically charged black hole is
\begin{equation}
  \varepsilon_{ex} = \frac{Q_m^2}{8\pi r_{_{H+}} }\left[1+ \frac{\alpha}{3\pi} \mathcal{K}_{NR+} \right].
\end{equation}
The properties of the maximally extractable energy in the strong
magnetic field case are similar to those of the magnetically charged
black hole in the weak magnetic field case. However, the QED
correction of the vacuum polarization effect to the energy of the
magnetically charged black hole in the strong magnetic field case is
much more significant in comparison with that in the weak magnetic
field case.

\section{Black holes with electric and magnetic charges}
\label{sec:NREMCBH}

If the spherically symmetric (nonrotating) black hole is both
electrically and magnetically charged, electric and magnetic fields
do not vanish. As shown in Eq.~(\ref{perturRS}), both invariants $S$
and $P$ contribute to the Euler-Heisenberg effective Lagrangian. The
metric takes the same form as the metric of Eq.~(\ref{anline}) for
electrically charged black holes with spherical symmetry. In this
case, the tensor $F_{\mu\nu}$ compatible with spherical symmetry can
involve only a radial electric field $F_{01} = - F_{10}$ and a
radial magnetic field $F_{23} = - F_{32}$, and the gauge potential
is (see, e.g., Ref.~\cite{Hawking1995})
\begin{equation} \label{gaugepem}
  A_\mu(x)= [A(r),0,0,Q_m (1-\cos{\theta})/4\pi].
\end{equation}
In the Einstein-Maxwell theory, $A(r) = -Q/(4\pi r)$, and the metric
function $f(r)$ of Eq.~(\ref{anline}) is given by (see, e.g.,
Ref.~\cite{Hawking1995})
\begin{equation} \label{metricfnremem}
  f(r) = 1 - \frac{2G M}{r} + \frac{GQ^2}{4\pi r^2} + \frac{GQ_m^2}{4\pi
  r^2}.
\end{equation}

In the Einstein-Euler-Heisenberg theory, we study the spherically
symmetric black hole with electric and magnetic charges in the
weak-field case. Using Eq.~(\ref{ide2}), we derive the
energy-momentum tensor with a radial electric field $E$ and a radial
magnetic field $B$,
\begin{equation} \label{emtnremcbhwf}
  T^{\mu\nu} = T_M^{\mu\nu} \left[ 1 + \frac{2\alpha }{45\pi E_c^2}
  (E^2 - B^2)
  \right] + g^{\mu\nu} \frac{\alpha}{90\pi E_c^2} \left[ (E^2 - B^2)^2 + 7 (E\cdot B)^2 \right] + \cdots.
\end{equation}
Analogous to the analysis of electrically/magnetically charged black
holes with spherical symmetry, we obtain the (0-0) component of
Einstein equations,
\begin{equation} \label{eeqnremcwf}
  \frac{2m'(r)}{r^2} = 4\pi \left[ E^2(r) + B^2(r) + \frac{\alpha}{15\pi}
  E^4(r)/E_c^2 - \frac{\alpha}{45\pi}
  B^4(r)/E_c^2 + \frac{\alpha}{9\pi E_c^2} E^2(r) B^2(r)\right].
\end{equation}
In addition, we obtain the field equations from
Eq.~(\ref{fequation}) (see also Ref.~\cite{Yajima2001}),
\begin{eqnarray}
  & & E(r) + \frac{2\alpha}{45\pi} E^3(r)/E_c^2 + \frac{\alpha B^2}{9\pi E_c^2} E(r) = \frac{Q}{4\pi
  r^2}, \label{eqenrem}\\
  & & B(r) = \frac{Q_m}{4\pi r^2}. \label{eqbnrem}
\end{eqnarray}
Note that the mixing terms of the electric and magnetic fields in
Eqs.~(\ref{eeqnremcwf}) and (\ref{eqenrem}) come from the
contribution of the invariant $P$. Introducing
$\overline{E}(r)\equiv E(r)/E_c$, we have
\begin{equation} \label{enrcmbh}
  \overline{E}(r) = E_Q - \frac{2\alpha}{45\pi} E^3_Q -
  \frac{\alpha}{9\pi} B_Q^2 E_Q + \cdots,
\end{equation}
up to the first order of $\alpha$. We substitute the solutions of
(\ref{eqbnrem}) and (\ref{enrcmbh}) into the Einstein equation
(\ref{eeqnremcwf}) and obtain the mass-energy function
\begin{equation} \label{eeqnremciwf}
  m(r) = M-\int_r^\infty 4\pi r^2 dr \frac{1}{2}E_c^2\left[ E_Q^2 + B_Q^2 - \frac{\alpha}{45\pi} E_Q^4
  - \frac{\alpha}{45\pi} B_Q^4 - \frac{\alpha}{9\pi} B_Q^2 E_Q^2 \right].
\end{equation}
Disregarding the QED correction of the Euler-Heisenberg effective
Lagrangian, Eq.~(\ref{eeqnremciwf}) gives the solution $m(r) =
M-Q^2/8\pi r-Q_m^2/8\pi r$ for the Reissner-Nordstr\"om black hole
with electric and magnetic charges. Performing the integration in
Eq.~(\ref{eeqnremciwf}), we approximately obtain
\begin{equation} \label{mnremcbhwf}
  m(r)= M - \frac{Q^2}{8\pi r}\left[1-
  \frac{\alpha}{225\pi}E_Q^2\right] - \frac{Q_m^2}{8\pi r}\left[1-
  \frac{\alpha}{225\pi}B_Q^2\right] + \frac{\alpha}{45\pi} \frac{Q_m^2}{8\pi r} E_Q^2.
\end{equation}
In the limit $Q\gg Q_m$, Eq.~(\ref{mnremcbhwf}) becomes
Eq.~(\ref{intma}) of the electrically charged black hole. On the
contrary, in the limit $Q_m \gg Q$, Eq.~(\ref{mnremcbhwf}) becomes
Eq.~(\ref{mnrmcbhwf}) of the magnetically charged black hole. In
order to study the effect of the $P$ term in the Euler-Heisenberg
effective Lagrangian, we consider the case with large $P$ and small
$S$, i.e., $Q_m \approx  Q$. In this situation,
Eq.~(\ref{mnremcbhwf}) becomes
\begin{equation} \label{mnremecbhwf}
  m(r)= M - \frac{Q^2}{8\pi r}\left[2-
  \frac{7\alpha}{225\pi}E_Q^2\right],
\end{equation}
for $Q_m = Q$, i.e., $S=0$ and large $P$. Comparing to the cases of
electrically/magnetically charged black holes, the QED correction to
the black hole energy becomes larger, which results from the
combination effects of the vacuum polarization on electric and
magnetic charges of black holes in the Einstein-Euler-Heisenberg
theory.

In the same way that has been discussed in previous sections, up to
the leading order of $\alpha$, we obtain the horizon radii
$r_{_{H+}}$ and $r_{_{H-}}$ from Eq.~(\ref{mnremecbhwf}),
\begin{eqnarray}
  r_{_{H+}} &=& GM + \sqrt{G^2 M^2 - \frac{G Q^2}{4\pi} \left[ 2 -
  \frac{7\alpha}{225\pi} E_{Q+}^2
  \right]}, \label{rpnremewf}\\
  r_{_{H-}} &=& GM - \sqrt{G^2 M^2 - \frac{G Q^2}{4\pi} \left[ 2 -
  \frac{7\alpha}{225\pi} E_{Q-}^2
  \right]}, \label{rnnremewf}
\end{eqnarray}
and the Christodoulou-Ruffini mass formula
\begin{equation} \label{crmnremewf}
  M = M_{ir} + \frac{Q^2}{16\pi GM_{ir}} \left[2-
  \frac{7\alpha}{225\pi}E_{Q+}^2\right],
\end{equation}
as well as the maximal energy extractable from a black hole
\begin{equation} \label{eenremecwf}
  \varepsilon_{ex} = \frac{Q^2}{8\pi r_{_{H+}}}\left[2-
  \frac{7\alpha}{225\pi}E_{Q+}^2\right].
\end{equation}
Analogously, we obtain the irreducible mass $M_{ir}$ by substituting
Eq.~(\ref{rpnremewf}) into Eq.~(\ref{mirnrwf}), and the black hole
entropy $S_{en}$ by substituting Eq.~(\ref{rpnremewf}) into
Eq.~(\ref{entropynrwf}). The irreducible mass $M_{ir}$, the black
hole entropy $S_{en}$, and the maximal energy extractable from a
black hole receive the same QED correction, but a factor of $7/2$
larger, as compared with their counterparts in the case of either
electrically or magnetically charged black holes in the weak-field
case.

\section{Summary}
\label{sec:sum}

In this article, in addition to the Maxwell Lagrangian, we consider
the contribution from the QED Euler-Heisenberg effective Lagrangian
to formulate the Einstein-Euler-Heisenberg theory. On the basis of
this theory, we study the horizon radius, area, total energy,
entropy, and irreducible mass as well as the maximally extractable
energy of spherically symmetric (nonrotating) black holes with
electric and magnetic charges. Our calculations are made up to the
leading order of the QED corrections in the limits of strong and
weak fields. Our results show that the QED correction of the vacuum
polarization results in the increase of the black hole horizon area,
entropy and irreducible mass, as well as the decrease of the black
hole total energy and maximally extractable energy. The reason is
that the QED vacuum polarization gives rise to the screening effect
on the black hole electric charge and the paramagnetic effect on the
black hole magnetic charge. The condition of the extremely charged
black hole $M=Q/4\pi$ or $M=Q_m/4\pi$ is modified [ see
Eqs.~(\ref{excondi}), (\ref{excondim}), and (\ref{excondimsf})],
which results from the screening and paramagnetic effects.

To end this article, we would like to mention that in the
Einstein-Euler-Heisenberg theory, it is worthwhile to study
Kerr-Newman black holes, whose electric field ${\bf E}$ and magnetic
field ${\bf B}$ are determined by the black hole mass $M$, charge
$Q$, and angular momentum $a$ \cite{Newman1965}. In addition, it
will be interesting to study the QED corrections in black hole
physics by taking into account the one-loop photon-graviton
amplitudes of the effective Lagrangian (\ref{DHaction})
\cite{Drummond1980} and its generalizations
\cite{Gilkey1975,Bastianelli2000,Barvinsky8590,Gusev2009,
Bastianelli2009}. We leave these studies for future work.

\vspace{0.5cm}
\noindent %
{\bf Acknowledgements}

Yuan-Bin Wu is supported by the Erasmus Mundus Joint Doctorate
Program by Grant Number 2011-1640 from the EACEA of the European
Commission. We thank the anonymous referee for drawing our attention
to the one-loop photon-graviton amplitudes.

\noindent %

\end{document}